\documentclass[prc,aps,twocolumn,floats,floatfix,showpacs,superscriptaddress]{revtex4}
\usepackage{graphicx}
\usepackage{longtable}
\begin{document}
\title{Near threshold electroproduction of the {$\omega$} meson at Q{${\mathbf{^2}} \approx$} 0.5 GeV{${\mathbf{^2}}$}}

\author{P. Ambrozewicz}
\altaffiliation{presently at Florida International University, Miami, Florida 33199}
\affiliation{Temple University, Philadelphia, Pennsylvania 19122}
\affiliation{Thomas Jefferson National Accelerator Facility, Newport News, Virginia 23606}

\author{J. Mitchell}
\affiliation{Thomas Jefferson National Accelerator Facility, Newport News, Virginia 23606}

\author{J. Dunne}
\altaffiliation{presently at Mississippi State University, Starkville, Mississippi 39762}
\affiliation{Thomas Jefferson National Accelerator Facility, Newport News, Virginia 23606}

\author{P. Markowitz}
\affiliation{Florida International University, Miami, Florida 33199}
\affiliation{Thomas Jefferson National Accelerator Facility, Newport News, Virginia 23606}

\author{C. J. Martoff}
\affiliation{Temple University, Philadelphia, Pennsylvania 19122}

\author{J. Reinhold}
\altaffiliation{presently at Florida International University, Miami, Florida 33199}
\affiliation{Argonne National Laboratory, Argonne, Illinois 60439}

\author{B. Zeidman}
\affiliation{Argonne National Laboratory, Argonne, Illinois 60439}

\author{D. J. Abbott}
\affiliation{Thomas Jefferson National Accelerator Facility, Newport News, Virginia 23606}

\author{A. Ahmidouch}
\altaffiliation{presently at North Carolina A \& T State University, Greensboro, North Carolina 27411}
\affiliation{Hampton University, Hampton, Virginia 23668}
\affiliation{Kent State University, Kent, Ohio 44242}

\author{C. S. Armstrong}
\affiliation{College of William and Mary, Williamsburg, Virginia 23187}

\author{J. Arrington}
\affiliation{Argonne National Laboratory, Argonne, Illinois 60439}

\author{K. A. Assamagan}
\affiliation{Brookhaven National Laboratory, Upton, New York 11973}
\affiliation{Hampton University, Hampton, Virginia 23668}

\author{K. Bailey}
\affiliation{Argonne National Laboratory, Argonne, Illinois 60439}

\author{O. K. Baker}
\affiliation{Hampton University, Hampton, Virginia 23668}
\affiliation{Thomas Jefferson National Accelerator Facility, Newport News, Virginia 23606}

\author{S. Beedoe}
\affiliation{North Carolina A \& T State University, Greensboro, North Carolina 27411}

\author{H. Breuer}
\affiliation{University of Maryland, College Park, Maryland 20742}

\author{R. Carlini}
\affiliation{Thomas Jefferson National Accelerator Facility, Newport News, Virginia 23606}

\author{J. Cha}
\affiliation{Hampton University, Hampton, Virginia 23668}

\author{G. Collins}
\affiliation{University of Maryland, College Park, Maryland 20742}

\author{C. Cothran}
\affiliation{University of Virginia, Charlottesville, Virginia}

\author{W. J. Cummings}
\affiliation{Argonne National Laboratory, Argonne, Illinois 60439}

\author{S. Danagoulian}
\affiliation{North Carolina A \& T State University, Greensboro, North Carolina 27411}
\affiliation{Thomas Jefferson National Accelerator Facility, Newport News, Virginia 23606}

\author{D. Day}
\affiliation{University of Virginia, Charlottesville, Virginia}

\author{F. Duncan}
\affiliation{University of Maryland, College Park, Maryland 20742}

\author{D. Dutta}
\altaffiliation{presently at Duke University, Durham, North Carolina 27708}
\affiliation{Northwestern University, Evanston, Illinois 60201}

\author{T. Eden}
\affiliation{Hampton University, Hampton, Virginia 23668}

\author{R. Ent}
\affiliation{Thomas Jefferson National Accelerator Facility, Newport News, Virginia 23606}

\author{L. Ewell}
\altaffiliation{presently at University of Michigan Medical Center, Ann Arbor, Michigan 48109}
\affiliation{University of Maryland, College Park, Maryland 20742}

\author{H. T. Fortune}
\affiliation{University of Pennsylvania, Philadelphia, Pennsylvania 19104}

\author{H. Gao}
\affiliation{Argonne National Laboratory, Argonne, Illinois 60439}

\author{D. F. Geesaman}
\affiliation{Argonne National Laboratory, Argonne, Illinois 60439}

\author{P. Gueye}
\affiliation{Hampton University, Hampton, Virginia 23668}

\author{K. K. Gustafsson}
\altaffiliation{presently at University of Helsinki, Helsinki, Finland}
\affiliation{University of Maryland, College Park, Maryland 20742}

\author{J.-O. Hansen}
\altaffiliation{presently at Thomas Jefferson National Accelerator Facility, Newport News, Virginia 23606}
\affiliation{Argonne National Laboratory, Argonne, Illinois 60439}

\author{W. Hinton}
\affiliation{Hampton University, Hampton, Virginia 23668}

\author{C. E. Keppel}
\affiliation{Hampton University, Hampton, Virginia 23668}
\affiliation{Thomas Jefferson National Accelerator Facility, Newport News, Virginia 23606}

\author{A. Klein}
\affiliation{Old Dominion University, Norfolk, Virginia }

\author{D. Koltenuk}
\altaffiliation{presently at MIT Lincoln Laboratory, Lexington, Massachusetts 02420}
\affiliation{University of Pennsylvania, Philadelphia, Pennsylvania 19104}

\author{D. J. Mack}
\affiliation{Thomas Jefferson National Accelerator Facility, Newport News, Virginia 23606}

\author{R. Madey}
\affiliation{Hampton University, Hampton, Virginia 23668}
\affiliation{Kent State University, Kent, Ohio 44242}

\author{D. G. Meekins}
\altaffiliation{presently at Thomas Jefferson National Accelerator Facility, Newport News, Virginia 23606}
\affiliation{College of William and Mary, Williamsburg, Virginia 23187}

\author{H. Mkrtchyan}
\affiliation{Yerevan Physics Institute, Yerevan, Armenia 375036}

\author{R. M. Mohring}
\altaffiliation{presently at Millennium Cell, Inc., Eatontown, NJ 07724}
\affiliation{University of Maryland, College Park, Maryland 20742}

\author{S. K. Mtingwa}
\affiliation{North Carolina A \& T State University, Greensboro, North Carolina 27411}

\author{G. Niculescu}
\altaffiliation{presently at James Madison University, Harrisonburg, Virginia 22807}
\affiliation{Hampton University, Hampton, Virginia 23668}

\author{I. Niculescu}
\affiliation{Hampton University, Hampton, Virginia 23668}

\author{T. G. O'Neill}
\altaffiliation{presently at Sun Microsystems, Inc. Mountain View, California 94043}
\affiliation{Argonne National Laboratory, Argonne, Illinois 60439}

\author{D. Potterveld}
\affiliation{Argonne National Laboratory, Argonne, Illinois 60439}

\author{J. W. Price}
\affiliation{Rensselaer Polytechnic Institute, Troy, New York 12180}

\author{B. A. Raue}
\affiliation{Florida International University, Miami, Florida 33199}

\author{P. Roos}
\affiliation{University of Maryland, College Park, Maryland 20742}

\author{G. Savage}
\affiliation{Hampton University, Hampton, Virginia 23668}

\author{R. Sawafta}
\affiliation{North Carolina A \& T State University, Greensboro, North Carolina 27411}
\affiliation{Thomas Jefferson National Accelerator Facility, Newport News, Virginia 23606}

\author{ R. E. Segel}
\affiliation{Northwestern University, Evanston, Illinois 60201}

\author{S. Stepanyan}
\affiliation{Yerevan Physics Institute, Yerevan, Armenia 375306}

\author{V. Tadevosyan}
\affiliation{Yerevan Physics Institute, Yerevan, Armenia 375306}

\author{L. Tang}
\affiliation{Hampton University, Hampton, Virginia 23668}
\affiliation{Thomas Jefferson National Accelerator Facility, Newport News, Virginia 23606}

\author{B. P. Terburg}
\affiliation{University of Illinois, Champaign-Urbana, Illinois 61801}

\author{S. Wood}
\affiliation{Thomas Jefferson National Accelerator Facility, Newport News, Virginia 23606}

\author{C. Yan}
\affiliation{Thomas Jefferson National Accelerator Facility, Newport News, Virginia 23606}

\author{B. Zihlmann}
\altaffiliation{presently at University of Gent, B-9000 Gent,  Belgium}
\affiliation{University of Virginia, Charlottesville, Virginia}

\date{\today}
\pacs{25.30.Rw, 25.30.Dh, 13.60.Le}

\begin{abstract}
Electroproduction of the $\omega$ meson was investigated in the $^1$H$(e,e^\prime p)\omega$ reaction.
The measurement was performed at a 4-momentum transfer $Q^2$ $\approx$ 0.5 GeV$^{\textrm{2}}$.
Angular distributions of the virtual photon-proton center-of-momentum cross sections have been extracted over the full angular range. 
These distributions exhibit a strong enhancement over $t$-channel parity exchange processes in the backward direction.
According  to a newly developed electroproduction model, this enhancement provides significant evidence of resonance formation 
in the $\gamma^* p \longrightarrow \omega p$ reaction channel.
\end{abstract}

\maketitle

\section{INTRODUCTION}

\begin{figure}[hb!]
\includegraphics[bb = 50 50 500 200, width=0.55\textwidth, keepaspectratio]{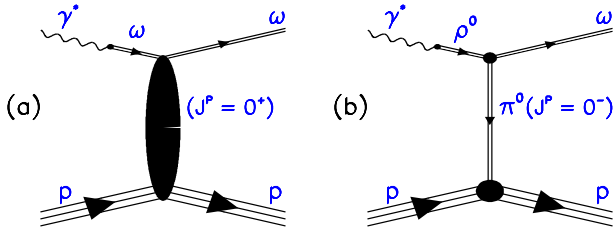}
\caption{\label{vmd}Vector meson dominance $t$-channel contributions: (a) diffractive scattering - natural parity exchange, 
(b) $\pi^0$ exchange - unnatural parity exchange.}
\end{figure}

There are only few measurements of the cross section for electroproduction of light vector mesons in the near threshold 
regime~\cite{joos76,joos77}. These experiments, carried out at DESY, despite suffering from very low statistics
revealed that different mechanisms contribute to production of the $\rho^0$ and $\omega$ mesons in this region. 
The data for both the energy dependence and angular distribution of $\rho^0$ meson 
electroproduction were found to be consistent with a vector meson dominance (VMD) model described by {\textit t}-channel
particle exchange with natural or unnatural parity. This production mechanism is represented by the {\textit t}-channel
diagrams of Fig.~\ref{vmd}. Diffractive scattering, interpreted as {\textit t}-channel Pomeron exchange in the language of Regge
theory, is the dominant process in the natural parity exchange mechanism above the traditional resonance region.
Near the $\omega$ production threshold, because of the appreciable relative decay width $\Gamma_{\omega \rightarrow \pi^0 \gamma}$ 
({$\sim\textrm{8}\%$}), {\textit t}-channel unnatural parity exchange, mediated by the exchange of the $\pi^0$ meson,
can make significant, even dominant, contributions to $\omega$ electroproduction. \\
\vspace{-0.1cm}
\indent A VMD-based model~\cite{fraas71}, which includes both of these mechanisms fails, however, to reproduce 
the electroproduction data near threshold~\cite{joos77}. It was found that the strength of the total cross section at threshold is much 
larger than that predicted for the $t$-channel exchange contributions. This enhancement was associated with the non-peripheral component 
of the total cross section corresponding to large $t$ or, equivalently, backward scattering angles.
Theoretical models based on $t$-channel exchange predict a strongly forward peaked angular distribution of the cross section that 
monotonically decreases with increasing angle. The results presented in this paper substantially differ from this prediction.
Such discrepancies were suggested by other earlier measurements which, as in Ref.~\cite{joos77}, found disagreements in the energy 
dependence of the total cross section~\cite{abbhhm68,klein96}. More recent theoretical models address this by including
$s$-channel and $u$-channel contributions to compensate for the additional strength at threshold.\\
\indent The data for the present analysis were acquired in Hall C at the Thomas Jefferson National Accelerator Facility (Jefferson Lab)
during an experiment designed to study electroproduction of strangeness via $^{1}{\textrm{H}}(e,e^\prime K^+)\Lambda(\Sigma)$~\cite{e91016}. 
Part of the background in the kaon electroproduction experiment were moderately inelastic $e^\prime p$ events rejected in the analysis by 
kaon particle identification. 
These $e^\prime p$ events, analyzed in the present work, provide the largest, to date, available data set on 
$\omega$ meson electroproduction.\\
\indent This work reports on a measurement of the differential cross section for electroproduction of $\omega$ mesons 
observed in the $^1{\textrm{H}}(e,e^\prime p)\omega$ reaction near threshold at four-momentum transfer
$Q^2 \approx 0.5$ GeV$^2$. The detailed analysis can be found in Ref.~\cite{pamb01}. 

\section{EXPERIMENT}

\begin{figure}[htb!]
\includegraphics[bb = 105 74 560 463, width=0.485\textwidth, keepaspectratio]{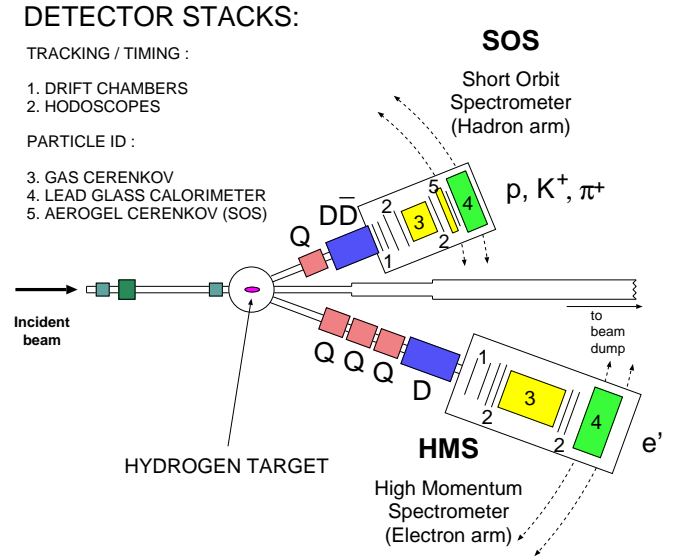}
\caption{\label{layout}Top view of Hall C.  {\textbf{Q}} and {\textbf{D}} denote quadrupole and dipole magnets respectively.}
\end{figure}

The experiment was conducted in Hall C at Jefferson Lab. The layout of the instrumentation is indicated in Fig.~\ref{layout}. 
Data were taken using 3.245 GeV electrons impinging on a 4.36-cm long target 
cell~{\cite{dunne97,meekins99}}. Liquid hydrogen circulating through the cell was cooled in a heat exchanger by 15 K gaseous 
helium and kept at a temperature of (19$\pm$0.2) K and a pressure of 24 psia.
\indent The experiment used the High Momentum Spectrometer (HMS) to detect scattered electrons. Its geometrical acceptance of 
$\sim 6.8$ msr was defined by an octagonal aperture in a 6.35-cm thick tungsten collimator. Before being detected,
the electrons traversed the magnetic field of four superconducting magnets; three quadrupoles followed by a dipole.
A pair of drift chambers at the focal plane of the spectrometer was used to determine the electron momentum while a threshold gas 
\v{C}erenkov detector and Pb-glass calorimeter provided particle identification at both hardware (trigger) and software levels. 
Arrays of segmented scintillator hodoscopes were used to form the trigger and provide  time-of-flight (ToF) measurements.
All of the $\omega$ data were taken with an HMS spectrometer central angle of 17.20$^\circ$ and a central momentum of 1.723 GeV.
This choice defined the virtual photon flux centered at 17.67$^\circ$ from the beam direction, and the four-momentum 
transfer $Q^2 \approx$ 0.5 GeV$^{\textrm{2}}$.\\

\begin{table}[h!]
\caption{\label{coverage} Central values of the hadron arm momentum $p_0$, angular setting $\theta_0$, as well as
the corresponding virtual photon proton separation $\theta_{\gamma p}$, and virtual photon $\omega$ meson
CM angle $\theta^*$.}
\begin{ruledtabular}
\begin{tabular}{cccc}
 \multicolumn{1}{c}{{$p_0$} (GeV)} & 
 \multicolumn{1}{c}{{$\theta_0$} (deg)} & 
 \multicolumn{1}{c}{{$\theta_{\gamma p}$} (deg)} &
 \multicolumn{1}{c}{{$\theta^*$} (deg)} \\
\hline
 1.077 & 17.67 & 0.00 & 180 \\ \cline{2-4}
       & 22.00 & 4.33 & 155 \\ \cline{2-4}
       & 26.50 & 8.78 & 135 \\ \cline{2-4}
       & 31.00 & 13.3 & 115 \\ \cline{1-4}
 0.929 & 17.67 & 0.00 & 180 \\ \cline{2-4}
       & 22.00 & 4.33 & 130 \\ \cline{2-4}
       & 26.50 & 8.78 & 110 \\ \cline{2-4}
       & 31.00 & 13.3 &  95 \\ \cline{2-4}
       & 35.00 & 17.3 &  85  \\ \cline{1-4}
 0.650 & 17.67 & 0.00 &   0   \\ \cline{2-4}
       & 22.00 & 4.33 &  15  \\ \cline{2-4}
       & 26.50 & 8.78 &  25  
\end{tabular}
\end{ruledtabular}
\end{table}
\indent The Short Orbit Spectrometer (SOS) was set to detect positively charged particles ({$\pi^+$, $K^+$} or {$p$}) and served as 
the hadron arm in the experiment. 
An octagonal aperture in a 6.35-cm thick tungsten collimator defined the SOS solid angle acceptance to be roughly 7.5 msr.
Hadrons were detected after passing through the magnetic field of three resistive magnets; a quadrupole and two dipoles
with opposite bending directions. 
A detector package similar to that of the HMS allowed for momentum determination (multi-wire drift chambers) and 
particle identification (segmented hodoscope arrays and \v{C}erenkov detectors).
Having fixed the electron arm position and momentum, the angular $\theta_0$ and momentum $p_0$ setting of the hadron arm was varied to access 
different scattering angles $\theta^*$ in the hadron ({$\gamma^* p$}) center-of-momentum (CM) system. 
These spectrometer settings, which corresponded to increasing virtual photon proton angular separation $\theta_{\gamma p}$ in the lab, allowed 
complete coverage for the $\omega$ scattering angles $\theta^*$ with respect to the virtual photon direction in the CM frame, particularly 
backward of 60$^o$. The data taken for the forward angles suffered from very low statistics. 
All the settings are presented in Table~\ref{coverage}.
\begin{figure}[htb!]
\includegraphics[bb = 0 0 850 567, width=0.5\textwidth, keepaspectratio]{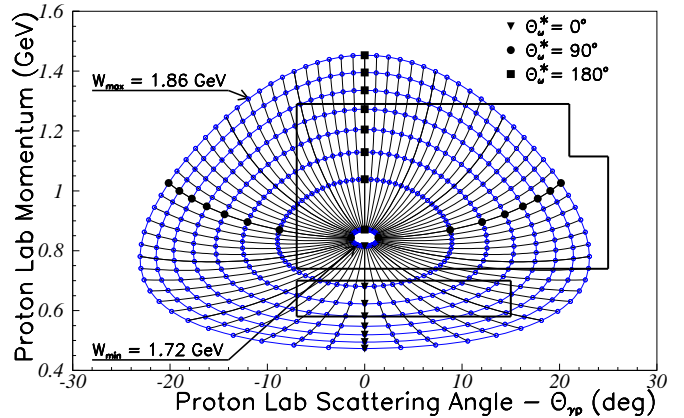}
\caption{\label{kincov}Total kinematic coverage. Straight lines define the acceptance of the experimental apparatus for 
all the kinematic settings.}
\end{figure}
\indent Figure~\ref{kincov} shows the full kinematic coverage of the data set in conjunction with the available acceptance. 
The closed curves in this figure are contours of constant invariant mass $W$ and the radial lines are contours of constant scattering 
angle $\theta^*$ in the hadron CM frame. Open circles are at 20 MeV and 5 degree increments, respectively.
The plot was generated for the $\omega$ mass, 0.782 GeV, and $Q^2 = 0.5$ GeV$^{{\textrm 2}}$. 
It is evident from this plot that a finite acceptance in proton lab momentum can produce cuts in which the range of accepted
$W$ is a strong function of $\theta^*$. 
These correlations were accounted for in the extraction of the differential cross sections from the data.

\section{DATA ANALYSIS}

\begin{figure}[ht!]
\includegraphics[bb = 0 0 950 567, width=0.5\textwidth, keepaspectratio]{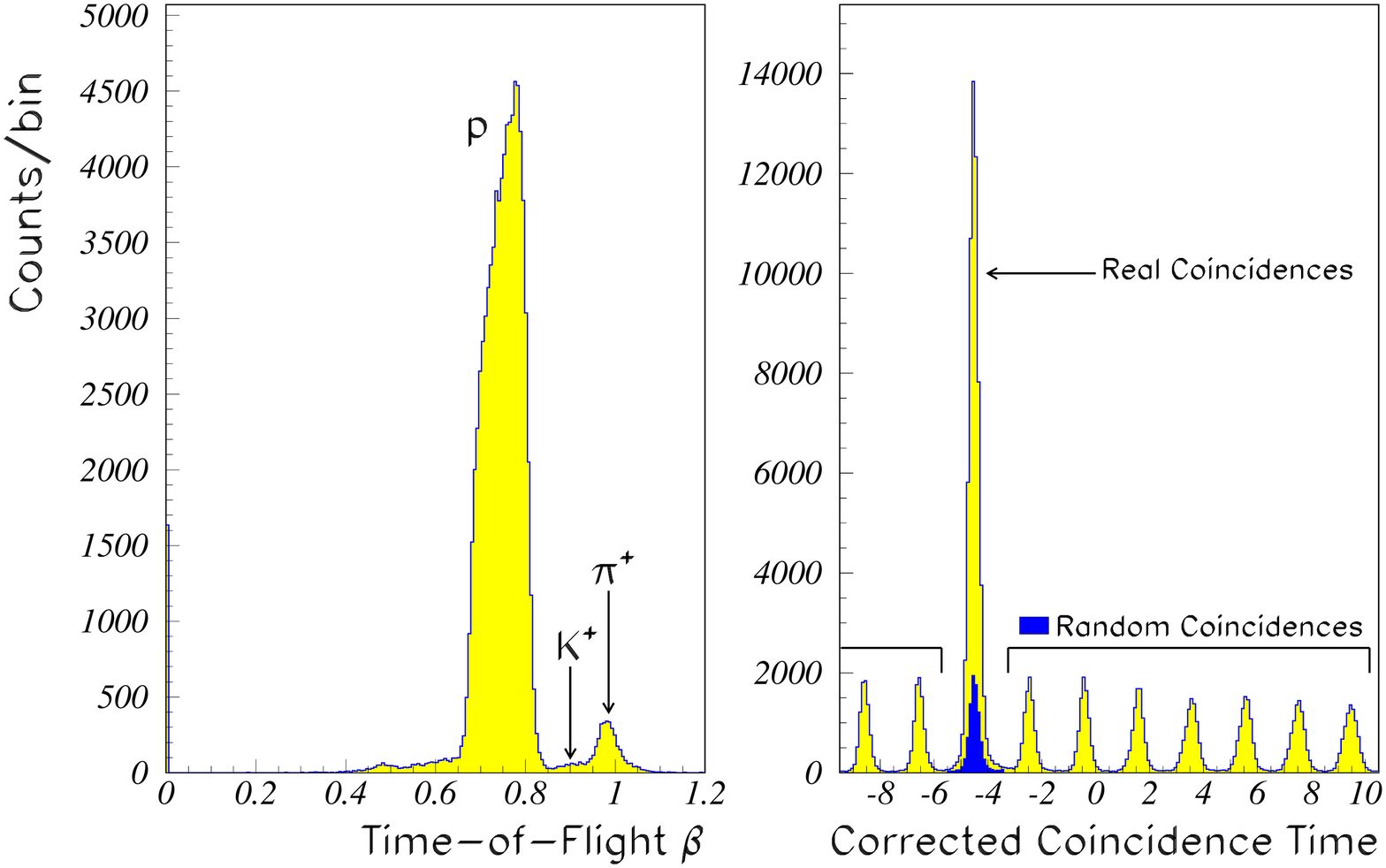}\\
\hspace{3mm}
\includegraphics[bb = 0 0 709 425, width=0.5\textwidth, keepaspectratio]{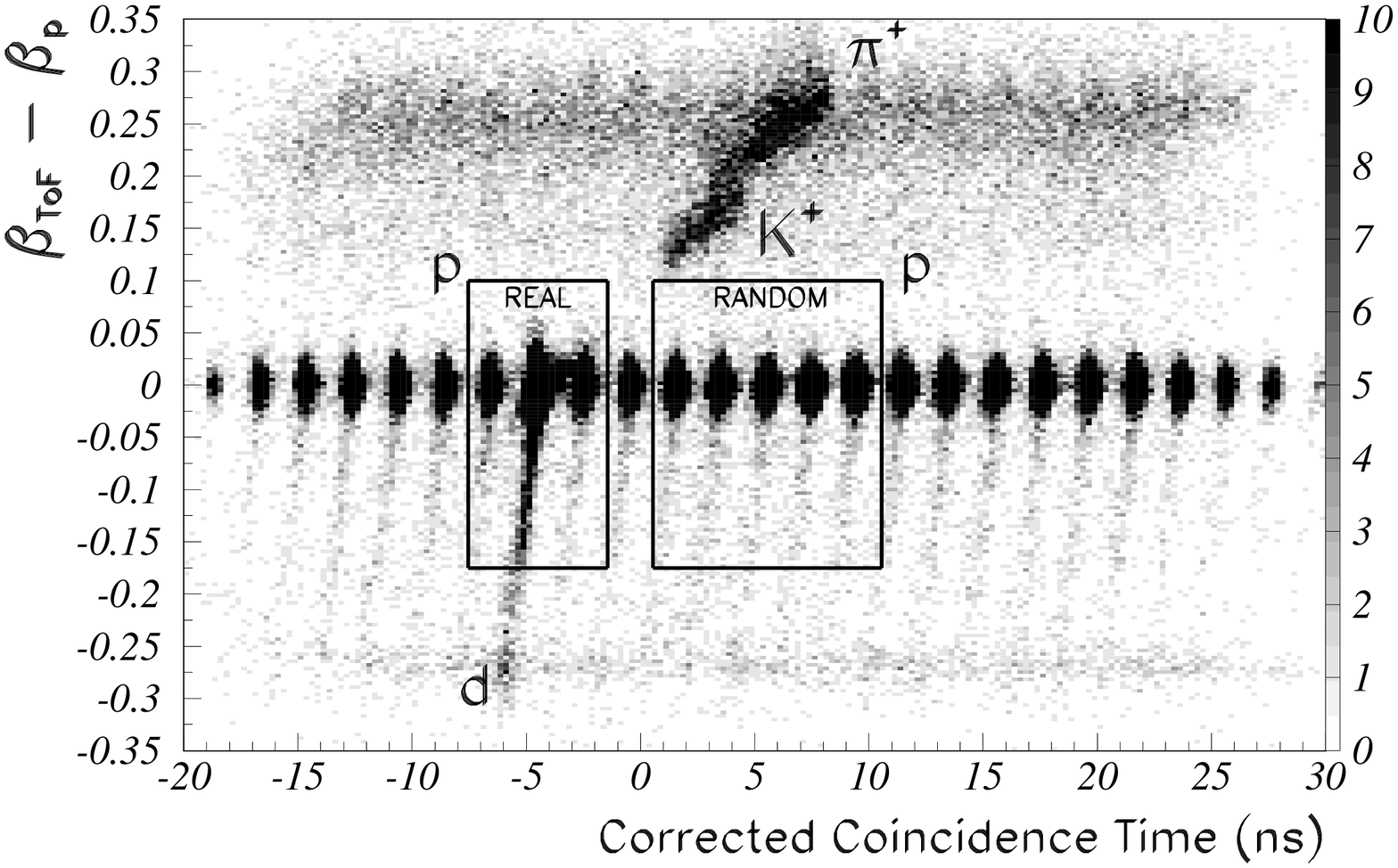}
\caption{\label{ppid}{{\textbf{\textrm{Top left}}}: Velocity distribution from the time-of-flight measurement 
({$\beta_{{\textrm{ToF}}}$}) for the real coincidence time cut shown in the bottom plot and described in the text. 
{\textbf{\textrm{Top right}}}: Distribution of the corrected coincidence time for protons. The estimated random coincidence contribution is 
overlayed on top of the coincident proton peak. Protons were selected using the ToF cut shown in the bottom plot.
{\textbf{\textrm{Bottom}}}: Typical spectrum of the difference in the velocities as determined by the time-of-flight technique and proton 
momentum versus the corrected coincidence time.}}
\end{figure}

\begin{figure}[hbt!]
\includegraphics[bb = 0 0 775 400, width=0.5\textwidth, keepaspectratio]{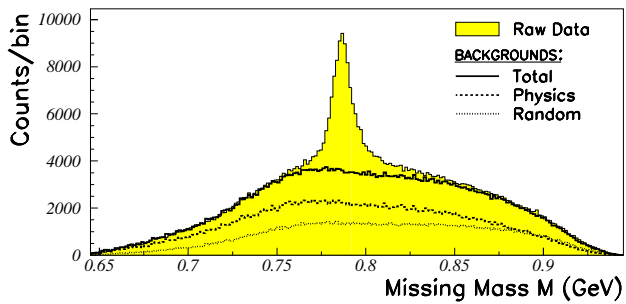}
\caption{\label{mxbgs}Missing mass distribution for $^1$H$(e,e^\prime p)X$ showing the decomposition into 
a peak for $\omega$ and the background.}
\end{figure}

Inelastic electron-proton final states were relatively easy to identify.
Electrons were well separated from pions at the trigger level and final purification was achieved by using cuts on detector responses 
from the HMS gas \^{C}erenkov detector and the Pb-glass calorimeter. Protons were selected using two types of scintillator timing 
information, time-of-flight (ToF) and coincidence time. In the SOS, the ToF was measured between two pairs of segmented hodoscope arrays
separated by $1.76$ m. In addition, relative coincidence time was measured between the hadron and electron arm scintillator arrays. 
The top plots in Fig.~\ref{ppid} show typical distributions of ToF velocity, $\beta_{{\textrm{ToF}}}$, and coincidence time. 
\indent The relatively large momentum acceptance, $\pm 20\%$ of the central setting (Table~\ref{coverage}), resulted in a variation of 
velocity with momentum (manifested as an asymmetry in the proton $\beta_{{\textrm{ToF}}}$ distribution, see Fig.~\ref{ppid} top left). This,
together with the associated pathlength variations, required corrections to the coincidence time to account for deviations
from the central trajectory.
The corrected coincidence time distribution (Fig.~\ref{ppid} top right) clearly shows the $2$ ns radio frequency (RF) 
microstructure of the electron beam. This structure was essential in the proton identification and accidental background removal. 
Real coincidence events, $e^\prime p$ pairs coming from the same interaction point, form a prominent peak at $-4.5$ ns. 
The remaining peaks are formed by random coincidences. \\
\indent The final sample of protons was selected by requiring the corrected coincidence time to be within the three RF peaks 
centered on the true coincidence peak and by employing a cut, for improved selectivity, on the difference between ToF velocity 
$\beta_{{\textrm{ToF}}}$ and the velocity calculated using the measured proton momentum $\beta_p$. This combination of cuts allowed 
the retention of those protons that underwent interactions in the SOS detector hut. These events form a shoulder that extends
from the proton coincident peak toward negative values of $\beta_{{\textrm{ToF}}}-\beta_p$ (Fig.~\ref{ppid} bottom). \\
\indent Random coincidences, also present beneath the true coincidence peak (Fig.~\ref{ppid} top right), contributed a background
in the final data sample (Fig.~\ref{mxbgs}). These were averaged and removed by selecting a sample of random coincidences from five 
RF peaks (the selection procedure is shown in the bottom of Fig.~\ref{ppid}). The random-subtracted distribution for any physics 
quantity was then obtained by subtracting the corresponding distribution for real and random samples, weighted by a 3:5 ratio to account 
for the differing numbers of peaks in the respective samples. \\
\begin{figure}[htb!]
\hspace{1.0cm}
\includegraphics[bb = 100 100 450 450, width=0.325\textwidth, keepaspectratio]{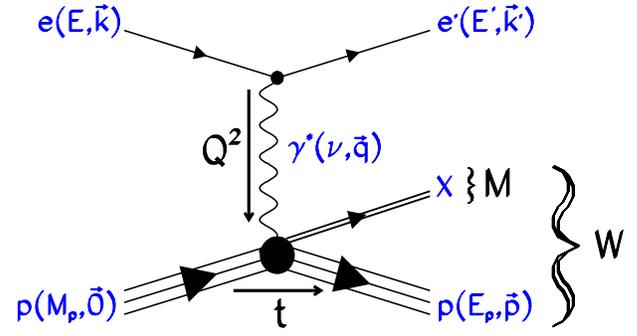}
\vspace{-0.5cm}
\caption{\label{bornterm}Fixed target {${\mathrm{^1H(e,e^\prime p)X}}$} scattering process.
Here, as well as in the text below, the energy and three-momentum transfer {$\nu$} and {$\vec{q}$} are given by 
{$\nu = E - E^\prime$} and {$\vec{q} = \vec{k} - \vec{k}^\prime$}.}
\end{figure}
\indent The kinematics of the $\omega$ channel for a fixed target is diagrammatically shown in Fig~\ref{bornterm}. 
Kinematic quantities characterizing the process can be expressed employing the notation of Fig.~\ref{bornterm}:
\begin{eqnarray}
Q^2   & = & - [(E,\vec{k}\;) - (E^\prime,\vec{k}^\prime)]^2  \stackrel{\;m_e \rightarrow 0\,}{\approx} 
4EE^\prime \sin^2\left(\theta_{e^\prime}/2\right)  \; ,\\
W^2   & = & [(M_p,\vec{0}\;) + (\nu,\vec{q}\;)]^2 = M_p^2 + 2M_p\nu - Q^2  \; ,\\
t\;\; & = & [(M_p,\vec{0}\;) - (E_p,\vec{p}\;)]^2 = 2M_p(M_p - E_p) \; , \label{t} \\
M^2 & = & [(M_p,\vec{0}\;) + (E,\vec{k}\;) - (E^\prime,\vec{k}^\prime) - (E_p,\vec{p}\;)]^2 = \nonumber \\ 
    & = & W^2 + M_p^2 - 2E_p(M_p + \nu) + 2|{\vec{q}}\,||{\vec{p}}\,|\cos \theta_{\gamma p} \; , \label{mx2}
\end{eqnarray}
where $\theta_{e^\prime}$ is the laboratory electron scattering angle and $\theta_{\gamma p}$ is the
proton scattering angle with respect to the virtual photon direction. $Q^2$ is square of the four-momentum transfer to the target, $W$ is
the invariant mass of the virtual photon-proton system, $t$ is the squared four-momentum transfer to the proton, and $M$ is
the mass of the system of undetected particles.
\begin{figure}[htb!]
\includegraphics[bb = 0 0 775 400, width=0.5\textwidth, keepaspectratio]{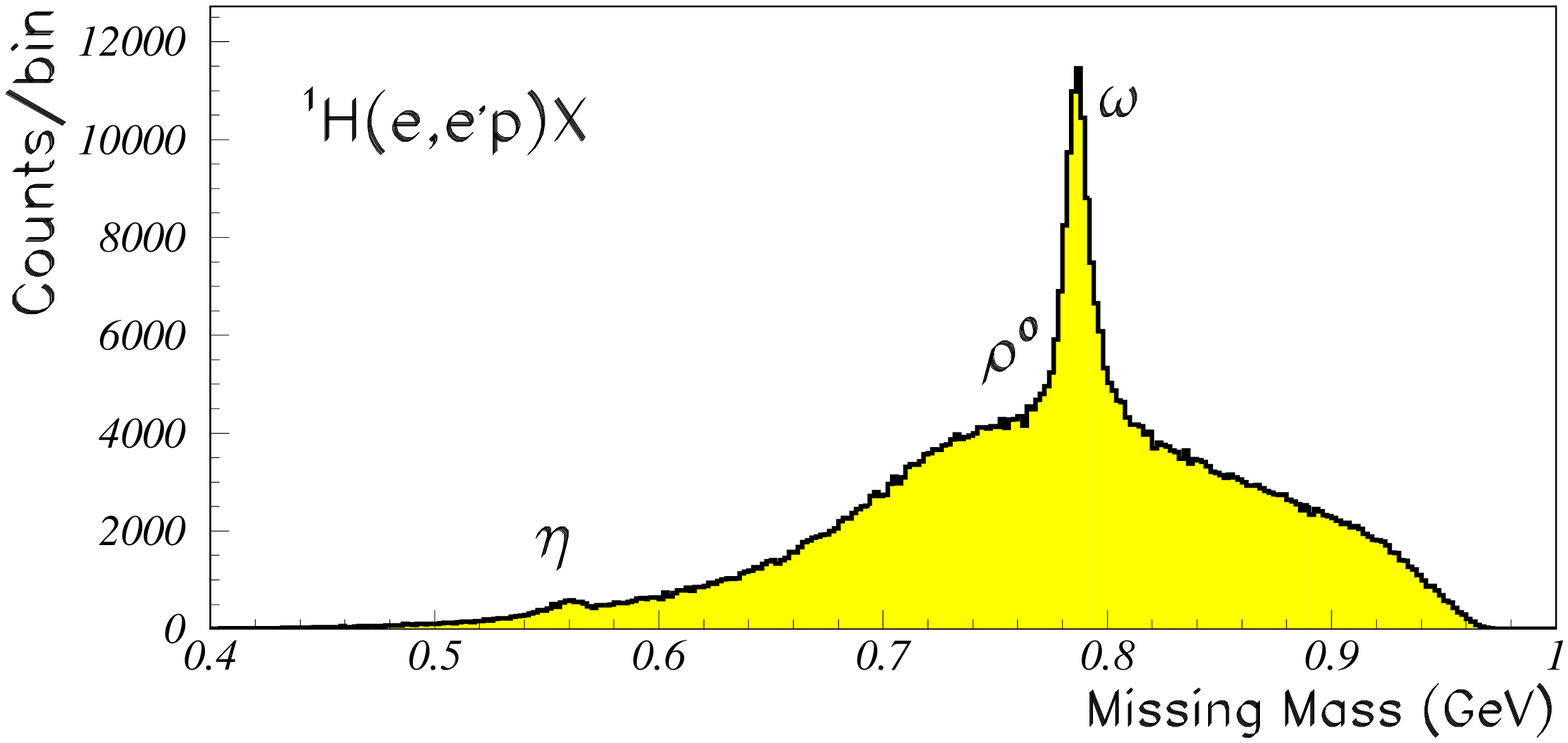}
\caption{\label{heep}Light vector meson electroproduction. The histogram shows events for all accepted momenta
for one setting. Note the presence of the pseudoscalar $\eta$ meson signal.}
\end{figure}
\indent Reconstruction of the missing mass, performed according to Eqn.~(\ref{mx2}), reveals a spectrum
with a strong $\omega$ meson signal atop a complicated background (Fig.~\ref{heep}). The data were corrected for trigger inefficiency 
($<\!{\mathrm{1}}\%$), track reconstruction inefficiencies ($\sim$10$\%$), particle ID inefficiencies ($\sim$2$\%$), 
and computer and electronic dead times ($\sim$5$\%$).
In the CM system, the virtual photon cross section for $\omega$ production ${d\sigma_v}/{d\Omega^*}$ is given in terms of 
the conventional two-particle coincidence cross section
\begin{equation}
\frac{d\sigma}{dp_{e^\prime} d\Omega_{e^\prime} d\Omega^*}  = \Gamma_T\;\frac{d\sigma_v}{d\Omega^*} \; ,
\end{equation} 
where $\Gamma_T$ is the virtual photon flux. The virtual photon cross section can be decomposed into transverse ({$\sigma_T$}), 
longitudinal ({$\sigma_L $}), and interference terms ({$\sigma_{TT}$}, {$\sigma_{LT}$}), such that
\begin{equation}\label{sigvdecomp}
\frac{d\sigma_v}{d\Omega^*} = 
\sigma_U + \varepsilon\,\cos 2\phi^*\,\sigma_{TT} + \sqrt{\frac{\varepsilon(\varepsilon+1)}{2}}\cos\phi^* \,\sigma_{LT} \; ,
\end{equation}
where $\sigma_U = \sigma_T + \varepsilon \, \sigma_L$, $\varepsilon$ is the virtual photon polarization parameter, and $\phi^*$ is 
the relative angle between the electron scattering plane and hadron production plane. \\

\indent The biggest challenge in cross section extraction was the separation of the data into the physics backgrounds and 
$\omega$ meson production (Fig.~\ref{mxbgs}). This was accomplished by using a Monte Carlo program to simulate both processes, the dominant 
background as well as $\omega$ production. The background was modeled as a combination of two processes, electroproduction of 
the neutral $\rho$ meson and multi-pion production.\\
\indent Production of the $\rho$ was assumed to be purely diffractive~{\cite{frasch69}},
\begin{eqnarray}\label{rhowgt}
\frac{d\sigma}{d\Omega^* dM} = \left(\frac{M_\rho}{M}\right)^n \; B_\rho(M) \;\; De^{bt^\prime} \; ,
\end{eqnarray}
where $t^\prime = t - t_{{\textrm{min}}}$, with $t_{{\textrm{min}}}$ being the momentum transfer when the scattering occurs 
along the virtual photon direction. In the above expression, coefficients $D$ and $b$ are $Q^2$ and $W$ dependent to account 
for their variation
near threshold and $D$, at $Q^2=0$, corresponds to the photoproduction cross section. 
The skewness of the $\rho$ meson shape, apparent from other experiments, was accounted for by using the Ross-Stodolsky 
parameterization~\cite{ross} (in Eqn.~(\ref{rhowgt}) first factor on the right-hand side) with the exponent $n=5.2$ coming from a fit 
to the DESY data~{\cite{joos76}}. 
For both the background and the $\omega$ meson, the mass distributions were generated according to a fixed width relativistic
Breit-Wigner distribution
\begin{equation}\label{lshape}
B_v(M) = \frac{M_v^2\Gamma_v^2}{(M^2 - M_v^2)^2 + M_v^2\Gamma_v^2} \; ,
\end{equation}
\noindent where $v$ is $\rho$ or $\omega$ with $M_\omega = 781.94$ MeV, $\Gamma_\omega = 8.43$  MeV, $M_\rho = 768.1$ MeV, 
and $\Gamma_\rho = 150.7$ MeV~{\cite{pdg98}}.

The multi-pion processes were collectively modeled as a Lorentz invariant electroproduction phase space for two-body production of 
a fictitious particle with arbitrary mass {\textit M}. This term is meant to account for all physically allowed reactions ({$W$} is well 
above the $\pi\pi$ threshold) that result in more than three particles (including the electron and proton) in the final state. The flatly 
distributed low yield of events (approximately $2\%$ at most settings) coming from the aluminum walls of the liquid hydrogen target were 
also treated as a part of the phase space background. The phase space was simulated by
\begin{eqnarray}\label{phspwgt}
\frac{d\sigma}{d\Omega^* dM} = \frac{1}{32\pi^2}\left(\frac{p^*}{q^*}\right)\frac{M}{W^2} \; ,
\end{eqnarray}
where $q^*$ and $p^*$ are the initial and final momenta in the CM frame, respectively. \\

\begin{figure}[ht!]
\includegraphics[bb = 0 0 585 709, width=0.5\textwidth, keepaspectratio]{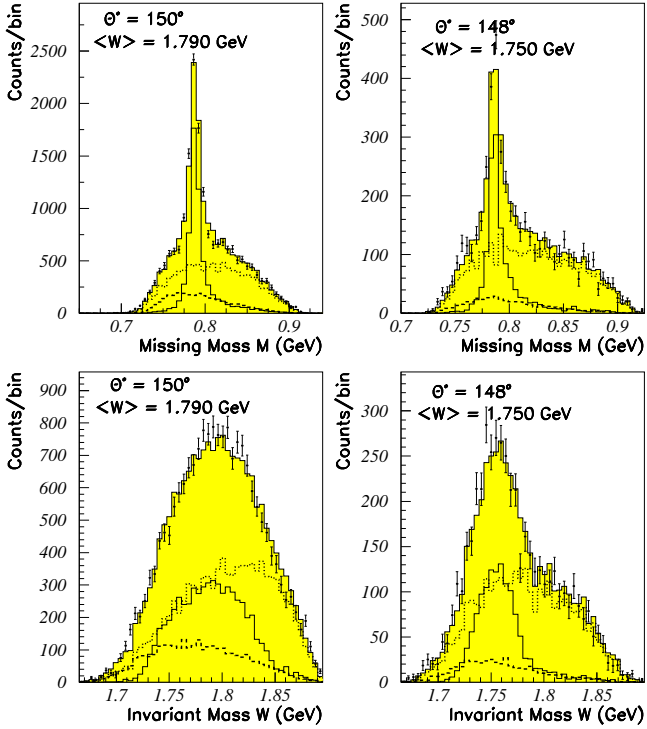}
\caption{\label{mxfit1}Missing mass {$M$} and invariant mass {$W$} distributions broken down into individual contributions for two different
momentum settings but the same angular settings. Solid circles with error bars are the data. The shaded histogram is 
the full Monte Carlo fit. The dotted line histogram corresponds to the resulting phase space yield, the solid line histogram 
to the {$\omega$} yield, and the dashed line histogram to the {$\rho^0$} yield.}
\end{figure} 
\indent The production of the $\omega$ meson was simulated with a cross section assumed to be $t$-channel unnatural parity exchange
\begin{eqnarray}\label{omegawgt}
\frac{d\sigma}{d\Omega^* dM} =
B_\omega(M) \; (\sigma_T^{\pi} + \varepsilon \, \sigma_L^{\pi}) \, ,
\end{eqnarray}

\noindent with $\sigma_T^{\pi}$, $\sigma_L^{\pi}$ being the transverse and longitudinal parts of the corresponding cross 
section~{\cite{fraas71}}. Within this model, the longitudinal contribution $\sigma_L^{\pi}$ is insignificant because it is
an order of magnitude smaller than $\sigma_T^{\pi}$ for the kinematic regime of the experiment. Natural parity exchange 
was neglected because it is also roughly one order of magnitude smaller than $\sigma_T^{\pi}$ within this regime. Similarly neglected 
were the nearly vanishing contributions from the interference terms $\sigma_{TT}$ and $\sigma_{LT}$. This amounts to modeling the total
cross section using only the largest contribution.
\begin{figure}[ht!]
\includegraphics[bb = 0 0 575 709, width=0.5\textwidth, keepaspectratio]{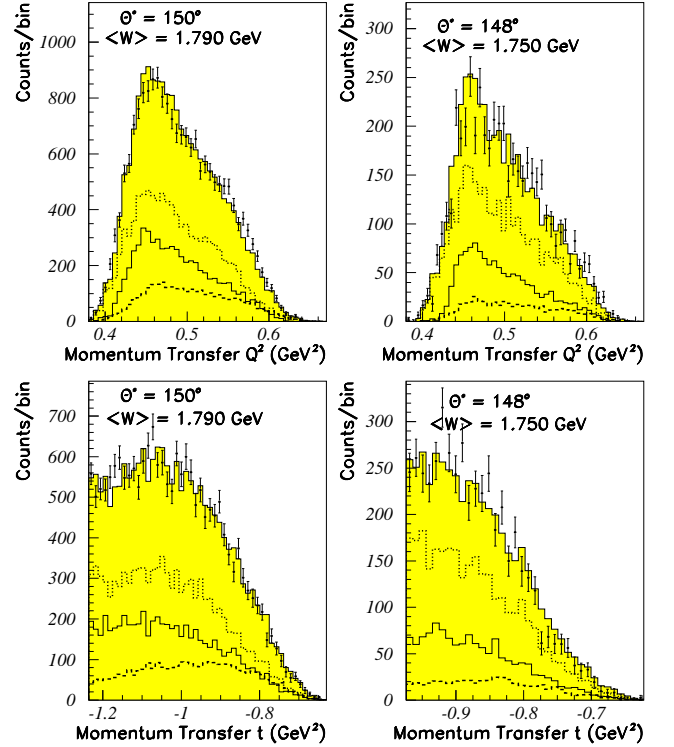}
\caption{\label{tfit1}Distributions of four-momentum transfers {$Q^2$} and {$t$} for the same settings as described in the caption 
of Fig.~\ref{mxfit1}. 
The cut-off at low values of {$-t$} reflects the proton momentum cut applied in the analysis since the proton energy is 
directly proportional to {$t$} in the fixed target regime (see Eqn.~\ref{t}).}
\end{figure}
The Monte Carlo program simulated finite target effects (multiple scattering and ionization energy losses), 
acceptance corrections, and radiative proccesses. The radiative corrections were modeled after the approximations
from Ref.~\cite{ent}. They were accounted for by altering the incident and scattered electron kinematics
and applying loop and vertex corrections which modify the cross section, but do not modify the missing mass distribution.
Having simulated all the processes for each kinematic setting, the data and Monte Carlo events were binned in CM scattering 
angle $\theta^*$.
Finally, a binned maximum likelihood fit was performed simultaneously in missing mass, $W$, $Q^2$, $t$, and $\theta^*$. 
The approach incorporated in the fit was developed by R. Barlow~{\cite{barlow93}}. The likelihood function accounted for fluctuations in 
the data and Monte Carlo distributions due to finite statistics. Its maximization allowed the search for the overall strengths, $p_i$, 
of each process modeled, so that the resulting yields for each bin satisfy the relation
\begin{equation}\label{prel}
Y_{\mathrm{DATA}} = Y_{\mathrm{MC}} = p_1\,Y_\omega + p_2\,Y_\rho + p_3\,Y_{\mathrm{phsp}} \; .
\end{equation}

\begin{figure}[hbt!]
\includegraphics[bb = 0 0 575 709, width=0.5\textwidth, keepaspectratio]{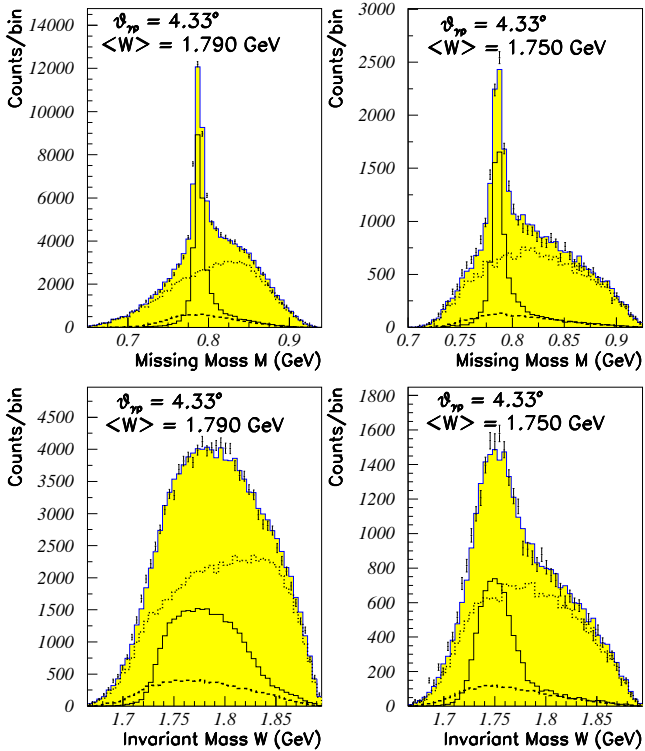}
\caption{\label{gfitmxw}Fits for two different momentum settings summed over $\theta^*$ bins for missing mass (top) and invariant mass 
(bottom). Fig.~\ref{mxfit1} contains the legend explanation.}
\end{figure}

\begin{figure}[hbt!]
\includegraphics[bb = 0 0 575 709, width=0.5\textwidth, keepaspectratio]{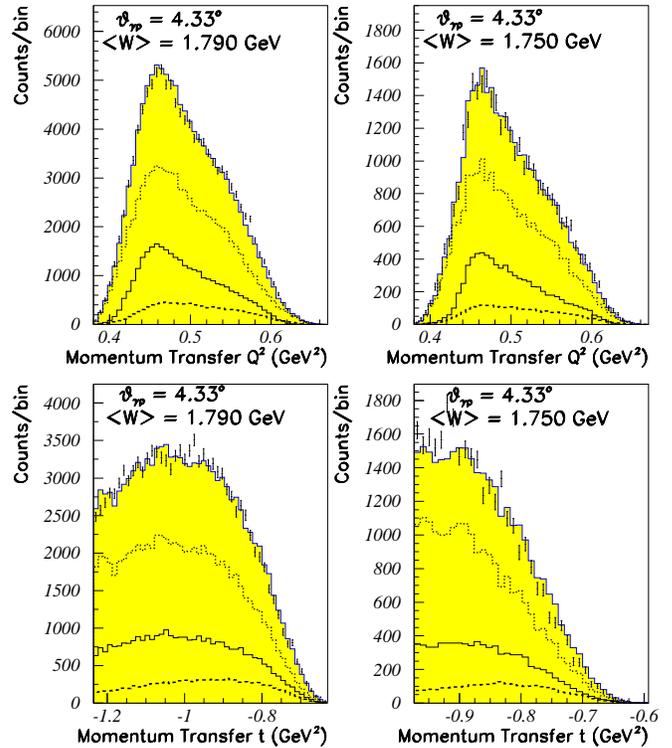}
\caption{\label{gfitq2t}Fits for two different momentum settings summed over $\theta^*$ bins for four-momentum transfer {$t$} (top) 
and four-momentum transfer {$Q^2$} (bottom). Fig.~\ref{mxfit1} contains the legend explanation.}
\end{figure}

\begin{figure}[ht!]
\includegraphics[bb = 0 0 575 709, width=0.5\textwidth, keepaspectratio]{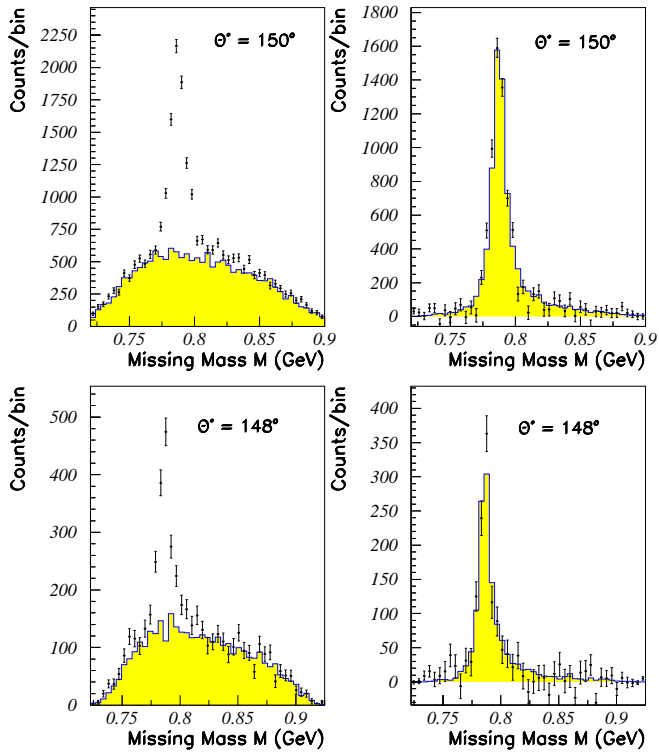}
\caption{\label{snlo}The results of signal-background separation, or, equivalently, $\omega$ yield extraction, 
for two different momentum settings of the hadron arm but the same, $\theta_{\gamma{\textrm{p}}}=4.33^o$, angular setting, 
Top: central momentum $p_{{\textrm 0}} = 1.077$ GeV.  Bottom: central momentum $p_{{\textrm 0}} = 0.929$ GeV.}
\end{figure}

Results of the fitting process for the high momentum setting of the hadron arm, {$p_0^{{\textrm{SOS}}} = 1.077$ }GeV, and 
the intermediate momentum setting, {$p_0^{{\textrm{SOS}}} = 0.929$ }GeV, for the same angular setting of 
{$\theta_{\gamma{\textrm {p}}} = 4.33^o$}, are shown in Fig.~\ref{mxfit1} and Fig.~\ref{tfit1}.
Figures~\ref{gfitmxw} and~\ref{gfitq2t} show the result of summation of the fits for all $\theta^*$ bins within
these two hadron arm settings, respectively, thus reflecting the goodness of the fit.
Performing the fit allowed separation of the raw data into the Monte Carlo determined background, consisting of 
the $\rho$ meson and phase-space contributions, and the $\omega$ meson signal, thus obtaining the data yields (Fig.~\ref{snlo}).
Subsequently, the differential virtual photon cross section was computed by scaling the model cross section by the data yield 
$ Y^{\omega}_{DATA}$ ($\approx p_1\,Y_{\omega}$), normalized to the simulated yield
\begin{eqnarray} \label{xsec}
\frac{d\sigma_v}{d\Omega^*} = \frac{Y^{\omega}_{DATA}}{Y^{\omega}_{MC}} \; \left(\frac{d\sigma}{d\Omega^*}\right)_{MC} \; .
\end{eqnarray} 
\noindent The Monte Carlo yield $Y^{\omega}_{MC}$ was evaluated by integrating the model cross section, $d\sigma$/$d\Omega^*dM$
over the entire acceptance of the apparatus and binning the result in the CM scattering angle. 
For any $\theta^*_i$ bin, this process can be expressed as
\begin{eqnarray}\label{ymc}
Y^{\omega}_{MC} = \int_{{\mathcal{A}}(\theta^*_i)}\Gamma_T\,{\mathcal{R}}\,\frac{d\sigma}{d\Omega^*dM}\,dQ^2dWd\phi_{e^\prime}d\Omega^*dM\;,
\end{eqnarray}
\noindent where ${\mathcal{R}}$ represents the multiplicative part of the radiative corrections and ${\mathcal{A}}(\theta^*_i)$ is 
the acceptance for the given $\theta^*_i$ bin. In the above expression, mass was integrated over the $\omega$ line shape 
(Eqn. \ref{lshape}). The cross section was extracted at $Q^2 \approx 0.5$ GeV$^{\textrm 2}$ for 74 bins in $\theta^{{\textrm*}}$, mostly for 
backward directions in the $\gamma^* p$ CM system. Here, the Hand~{\cite{hand63}} convention was adopted in evaluating the virtual photon 
flux $\Gamma_T$. Identifying the $\omega$ meson production 
using only the $e^\prime p$ final states introduced a statistical error of less than $25\%$. Systematic 
uncertainties associated with the background subtraction are less than $12\%$. Fixed electron kinematics and limited out-of-plane acceptance
reduced the range of accepted $\phi^*$ angles to $\pm 30^{\textrm{o}}$ about $0^{\textrm{o}}$ for the outermost angular setting 
($\theta_{\gamma {\textrm p}} = 17.3^o$). This $\phi$ cut was also applied to the data of all other settings.

\section{RESULTS}

With the use of the procedures described above, angular distributions of the differential 
cross sections for electroproduction of the $\omega$ meson were extracted for two different average values of the invariant mass $W$.
The data were divided into two sets according to the average $W$ which, for each data point, was determined using the results of the fit.
These two sets form the angular distributions that correspond to mean invariant masses $\langle W \rangle$ of 1.750 GeV and 1.790 GeV. 
The results are presented in Tables~\ref{wlo} and~\ref{whi}.\\
\begin{table}[htp!]
\begin{center}
{\textbf {\caption{Differential cross sections for the lower average $W$ (${\langle W \rangle} = 1.75$ GeV). The bin width is
5$^0$, centered on the quoted value, except for the first bin whose width is 10$^0$. The cross sections were extracted
for $|\phi^*| < 30^o$.\label{wlo}}}}
\hspace{-0.15cm}
\begin{ruledtabular}
{\scriptsize{
\begin{tabular}{cccccc}
 \multicolumn{1}{c} {$\theta^*$} &
 \multicolumn{1}{c} {$d\sigma/d\Omega^*$} & 
 \multicolumn{2}{c} {Uncertainty} & 
 \multicolumn{1}{c} {${{{\langle W \rangle}}}$} &
 \multicolumn{1}{c} {${{{\langle Q^2 \rangle}}}$} \\\cline{3-4}
 \multicolumn{1}{c} {(deg)} &
 \multicolumn{1}{c} {($\mu b/sr$)} & 
 \multicolumn{1}{c} {Stat.} & 
 \multicolumn{1}{c} {Syst.} & 
 \multicolumn{1}{c} {(GeV)} & 
 \multicolumn{1}{c} {$(GeV^2)$} \\
\hline 
   45  &  0.257  &  0.057  &  0.015  &  1.753  &  0.501 \\
   75  &  0.116  &  0.026  &  0.011  &  1.745  &  0.512 \\
   80  &  0.170  &  0.026  &  0.006  &  1.747  &  0.511 \\
   85  &  0.112  &  0.024  &  0.006  &  1.747  &  0.510 \\
   90  &  0.131  &  0.024  &  0.006  &  1.747  &  0.510 \\
   95  &  0.163  &  0.024  &  0.008  &  1.749  &  0.510 \\
  100  &  0.176  &  0.023  &  0.010  &  1.752  &  0.509 \\
  101  &  0.170  &  0.028  &  0.012  &  1.752  &  0.509 \\
  105  &  0.260  &  0.023  &  0.012  &  1.755  &  0.505 \\
  106  &  0.267  &  0.028  &  0.012  &  1.756  &  0.508 \\
  110  &  0.292  &  0.024  &  0.012  &  1.758  &  0.504 \\
  111  &  0.311  &  0.025  &  0.013  &  1.761  &  0.505 \\
  115  &  0.440  &  0.026  &  0.014  &  1.763  &  0.501 \\
  120  &  0.466  &  0.025  &  0.013  &  1.766  &  0.499 \\
  125  &  0.425  &  0.025  &  0.013  &  1.766  &  0.498 \\
  130  &  0.399  &  0.026  &  0.012  &  1.762  &  0.498 \\
  135  &  0.412  &  0.031  &  0.012  &  1.759  &  0.500 \\
  138  &  0.400  &  0.031  &  0.012  &  1.749  &  0.497 \\
  140  &  0.458  &  0.044  &  0.012  &  1.755  &  0.501 \\
  143  &  0.466  &  0.033  &  0.012  &  1.751  &  0.498 \\
  148  &  0.367  &  0.028  &  0.010  &  1.751  &  0.497 \\
  153  &  0.352  &  0.030  &  0.010  &  1.750  &  0.501 \\
  158  &  0.308  &  0.031  &  0.010  &  1.748  &  0.501 \\
  163  &  0.353  &  0.039  &  0.009  &  1.747  &  0.501 \\
  168  &  0.288  &  0.040  &  0.010  &  1.745  &  0.504 \\
  173  &  0.199  &  0.050  &  0.012  &  1.742  &  0.510 \\
\end{tabular}}}
\end{ruledtabular}
\end{center}
\vspace{-0.5cm}
\end{table}

\begin{table}[hp!]
\begin{center}
{\textbf {\caption{Differential cross sections for the higher average $W$ (${\langle W \rangle} = 1.790$ GeV). The bin width is
5$^0$, centered on the quoted value, except for the first two bins whose width is 10$^0$. The cross sections were extracted
for $|\phi^*| < 30^o$.\label{whi}}}}
\hspace{-0.15cm}
\begin{ruledtabular}
{\scriptsize{
\begin{tabular}{cccccc}
 \multicolumn{1}{c} {$\theta^*$} &
 \multicolumn{1}{c} {$d\sigma/d\Omega^*$} & 
 \multicolumn{2}{c} {Uncertainty} & 
 \multicolumn{1}{c} {${\langle W \rangle}$} &
 \multicolumn{1}{c} {${\langle Q^2 \rangle}$} \\\cline{3-4}
 \multicolumn{1}{c} {(deg)} &
 \multicolumn{1}{c} {($\mu b/sr$)} & 
 \multicolumn{1}{c} {Stat.} & 
 \multicolumn{1}{c} {Syst.} & 
 \multicolumn{1}{c} {(GeV)} & 
 \multicolumn{1}{c} {$(GeV^2)$} \\
\hline
   25  &  0.501  &  0.058  &  0.015  &  1.778  &  0.505 \\
   35  &  0.360  &  0.053  &  0.015  &  1.765  &  0.502 \\
   62  &  0.229  &  0.034  &  0.015  &  1.808  &  0.493 \\
   67  &  0.263  &  0.030  &  0.014  &  1.808  &  0.489 \\
   72  &  0.186  &  0.027  &  0.014  &  1.811  &  0.488 \\
   73  &  0.171  &  0.033  &  0.014  &  1.771  &  0.504 \\
   77  &  0.193  &  0.025  &  0.014  &  1.814  &  0.484 \\
   78  &  0.168  &  0.032  &  0.014  &  1.773  &  0.503 \\
   82  &  0.141  &  0.023  &  0.013  &  1.819  &  0.479 \\
   83  &  0.175  &  0.030  &  0.013  &  1.775  &  0.502 \\
   84  &  0.173  &  0.030  &  0.012  &  1.780  &  0.512 \\
   88  &  0.225  &  0.029  &  0.012  &  1.779  &  0.500 \\
   87  &  0.226  &  0.024  &  0.012  &  1.821  &  0.477 \\
   89  &  0.256  &  0.024  &  0.012  &  1.782  &  0.502 \\
   92  &  0.251  &  0.027  &  0.012  &  1.825  &  0.475 \\
   93  &  0.249  &  0.028  &  0.012  &  1.784  &  0.496 \\
   94  &  0.237  &  0.020  &  0.012  &  1.789  &  0.498 \\
   97  &  0.282  &  0.031  &  0.012  &  1.827  &  0.473 \\
   98  &  0.329  &  0.027  &  0.012  &  1.791  &  0.493 \\
   99  &  0.321  &  0.019  &  0.012  &  1.800  &  0.491 \\
  102  &  0.309  &  0.047  &  0.012  &  1.827  &  0.470 \\
  103  &  0.395  &  0.028  &  0.011  &  1.798  &  0.489 \\
  104  &  0.352  &  0.018  &  0.012  &  1.808  &  0.485 \\
  108  &  0.365  &  0.030  &  0.012  &  1.796  &  0.488 \\
  109  &  0.391  &  0.020  &  0.011  &  1.811  &  0.484 \\
  113  &  0.318  &  0.037  &  0.012  &  1.791  &  0.488 \\
  114  &  0.450  &  0.023  &  0.011  &  1.814  &  0.481 \\
  116  &  0.396  &  0.025  &  0.011  &  1.772  &  0.500 \\
  119  &  0.515  &  0.029  &  0.011  &  1.816  &  0.481 \\
  121  &  0.486  &  0.024  &  0.010  &  1.785  &  0.493 \\
  124  &  0.524  &  0.044  &  0.010  &  1.816  &  0.482 \\
  126  &  0.503  &  0.023  &  0.010  &  1.795  &  0.488 \\
  131  &  0.506  &  0.024  &  0.010  &  1.796  &  0.488 \\
  136  &  0.518  &  0.027  &  0.010  &  1.792  &  0.490 \\
  140  &  0.530  &  0.024  &  0.010  &  1.772  &  0.489 \\
  141  &  0.495  &  0.031  &  0.010  &  1.788  &  0.492 \\
  145  &  0.538  &  0.020  &  0.010  &  1.781  &  0.488 \\
  146  &  0.536  &  0.046  &  0.010  &  1.779  &  0.501 \\
  150  &  0.492  &  0.019  &  0.010  &  1.783  &  0.489 \\
  151  &  0.471  &  0.102  &  0.014  &  1.765  &  0.516 \\
  155  &  0.431  &  0.019  &  0.010  &  1.780  &  0.492 \\
  160  &  0.425  &  0.021  &  0.010  &  1.777  &  0.494 \\
  165  &  0.439  &  0.026  &  0.010  &  1.773  &  0.498 \\
  166  &  0.652  &  0.093  &  0.014  &  1.785  &  0.470 \\
  170  &  0.442  &  0.036  &  0.011  &  1.764  &  0.504 \\
  171  &  0.419  &  0.059  &  0.012  &  1.780  &  0.482 \\
  175  &  0.485  &  0.070  &  0.012  &  1.755  &  0.513 \\
  176  &  0.392  &  0.077  &  0.014  &  1.778  &  0.489 \\
\end{tabular}}}
\end{ruledtabular}
\end{center}
\end{table}

\indent These two sets of the data, however, do not constitute two independent angular distributions. There are large overlaps in the 
$W$ ranges for both distributions that can readily be seen in the bottom of Figs.~\ref{mxfit1} and~\ref{gfitmxw}.
Therefore, the cross sections of both angular distributions were scaled to a reference $W$ of 1.785 GeV.
This was done by rescaling their corresponding kinematic parts, {\textit i.e.} phase space factors normalized to the incoming particle flux.
Due to a significant variation with mass, the scaling factor was determined on an event-by-event basis 
and then averaged. The scaling can quantitatively be described by
\begin{equation}\label{scale}
\left(\frac{d\sigma}{d\Omega^*}\right)_{{\textrm{scaled}}} = 
\frac{\Gamma(W_{\textrm{ref}})}{\Gamma(W)}
 \; \frac{d\sigma}{d\Omega^*} \; ,
\end{equation}
\noindent where $\Gamma(W) = p^*(W)/q^*(W)W^2$ is a normalized phase space factor (compare with Eqn.~(\ref{phspwgt})) and $p^*$ and $q^*$ 
are, respectively, the 3-momenta in the CM frame of the $\omega$ and the virtual photon which, for fixed $W$, 
are determined only by the masses of the interacting particles.
The result of this procedure is shown in Fig.~\ref{results}. 
Correcting for the phase space, opening up above the threshold, removes practically all of the observed $W$ dependence.
It also shows that the shape of the distribution is not trivially induced by $W$ variations of the phase space factors.\\
\indent The enhancement of the backward-angle cross section over $t$-channel unnatural parity exchange (Fraas model, dashed line) 
is evident. This was suggested by the earlier electroproduction~{\cite{joos77}} and photoproduction~{\cite{abbhhm68,klein96}} data.
Such a departure from the smooth fall-off of the $t$-channel processes, either in the angular distribution or $t$-dependence, 
has been attributed, theoretically, to $s$- and $u$-channel resonance contributions. 
Even though the energy dependence may not be sensitive to the details of the model, since it is integrated over full angular range,
the inclusion of resonance formation was also necessary to reproduce the near threshold strength of the photoproduction 
cross section (see~{\cite{zhao00,zhao01}).
\begin{figure}[hb!]
\includegraphics[bb = 0 0 575 425, width=0.5\textwidth, keepaspectratio]{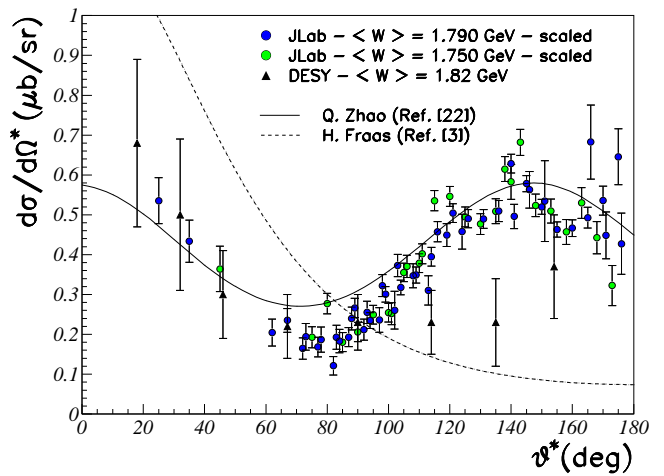}
\caption{\label{results}The angular distributions for different average {$W$} and for $|\phi^*| < 30^o$. Error bars are statistical. The DESY data correspond to the four-momentum transfer $Q^{\textrm{2}}=0.77$ GeV$^{\textrm{2}}$, $W = 1.82$ GeV and full $\phi^*$ 
range. The Fraas model shown here was used in the cross section extraction. Both distributions were scaled to $W=1.785$ GeV. 
The scaling procedure is described in the text.} 
\end{figure}
Recent examples of such calculations~{\cite{zhao00,zhao01,zlb98,yoh00a,yoh00b,zhao98}} mainly address SAPHIR data~{\cite{klein96}}.
Some of these works~{\cite{yoh00a,yoh00b}} showed that the dominant contributions could come from the missing resonances, $N_{3/2}^+(1910)$, 
and the $N_{3/2}^-(1960)$ (the latter is labeled $D_{13}(2080)$ by the Particle Data Group).
Other calculations, however, differ in predicting which nucleonic excitations could contribute in the $s$-channel. 
It was found that the contribution from two resonances, $P_{\textrm13}$(1720) and $F_{\textrm15}$(1680), dominated and their inclusion was 
necessary to reproduce the available photoproduction data near threshold~{\cite{zhao00,zhao01,zlb98,zhao98}}.\\ 
\indent From the point of view of the present work, the most interesting result of these theoretical models is that the nucleon 
resonances are the favored mechanism for producing backward-angle enhancements in the differential cross section.
The solid line in Fig.~\ref{results} shows the comparison of the data with an unpublished, as of this writing, electroproduction 
calculation~{\cite{zhao03}} complementary to the photoproduction model~\cite{zhao01}. 
In this model, the diffractive nature of $\omega$ production is described by Pomeron exchange based on Regge phenomenology and SU(3)
flavor symmetry. This contribution dominates the cross section above the resonance region. Neutral $\pi$ exchange in the $t$-channel
is included to account for the peaking of the cross section in the forward direction, especially near threshold. 
Resonance formation processes in the $s$- and $u$-channel that dominate intermediate and 
backward scattering angles, where the other contributions are small, were modeled in an $SU(6) \times O(3)$ quark model symmetry 
limit. All contributions, summed coherently, give a strongly $\phi^*$-dependent cross section (Eqn.~(\ref{sigvdecomp})). 
To correctly compare this theoretical calculation with the data,
the model was integrated over a range of the azimuthal angle $\phi^*$ corresponding to the cut used in the data analysis. 
The model was also averaged over the appropriate $W$ and $Q^{\textrm{2}}$ ranges.

\section{CONCLUSIONS}

\indent Cross sections for the $\omega$ meson electroproduction were obtained from the $^1$H$(e,e^\prime p)\omega$ reaction
at $E_e = 3.245$ GeV. The angular distribution of the differential cross section in the threshold regime has
unprecedented granularity and much smaller statistical uncertainties than in previous work. 
The angular distribution exhibits a substantial backward-angle 
enhancement of the cross section over the pure $t$-channel expectation, similar to that found in the DESY~{\cite{joos77}}, 
photoproduction~{\cite{abbhhm68}}, and SAPHIR data~{\cite{klein96}}.
 
\indent In comparing the result of this work to the Zhao model~\cite{zhao03},
the similarity of the angular distributions is evident. 
In the view of these results, this analysis provides significant evidence for resonance formation, 
possibly $s$-channel, in the $\gamma^\star p \longrightarrow \omega p$ reaction.
It is worth noting that, although elastic $\pi N$ scattering constitutes the main source of information on the nucleon excitation 
spectrum, it alone cannot distinguish among existing theoretical models~{\cite{isgur80}},
many of which predict a much richer baryonic, hence nucleonic, spectrum than currently 
observed~{\cite{isgur77,isgur78,karl78,karl79,capstick86,capstick94,capps74,cutkosky77,cutkosky83}}. 
If they exist, these states are either being masked by neighboring resonances with stronger couplings
or they are altogether decoupled from the $\pi N$ channel. There are decay modes, other than $\pi N$, however, that have sizeable resonance 
coupling constants~{\cite{capstick94,capstick92}}. A calculation, based on the symmetric quark model~{\cite{capsrob3}},
indeed predicts that vector meson decay channels, $N\rho$ and $N\omega$, have appreciable resonance couplings.
Electroproduction of $\omega$ mesons, enhanced by its isospin selectivity, may therefore provide additional evidence 
in the search for resonances unobserved in $\pi N$ scattering.

\begin{acknowledgments}
The authors would like to express sincere thanks to Dr. Qiang Zhao for sharing his electroproduction calculation and
for fruitful discussions on the underlying theory.
The authors would like to acknowledge the support of the staff of the Accelerator division of Jefferson Lab. This work was supported 
in part by the U.S. Department of Energy under contract W-31-109-Eng-38 for Argonne National
Laboratory, by contract DE-AC05-84ER40150, under which the Southeastern Universities Research Association (SURA) operates 
the Thomas Jefferson National Accelerator, and the National Science Foundation (grant no. NPS-PHY-9319984). It was also in part 
supported by Temple University, Philadelphia PA. \\
\indent Pawel Ambrozewicz would like to thank Dr. Kees de Jager, the leader of Hall A
at Jefferson Lab, for the support that allowed him to finish the analysis presented in this paper.
\end{acknowledgments}

\bibliographystyle{apsrev}
\bibliography{omega}

\end{document}